\documentstyle[12pt,epsfig,graphics,cite,amssymb,amsmath,bm,cool]{article}
\begin{document}\bibliographystyle{plain}\begin{titlepage}
\renewcommand{\thefootnote}{\fnsymbol{footnote}}\hfill
\begin{tabular}{l}HEPHY-PUB 938/14\\UWThPh-2014-19\\August
2014\end{tabular}\\[2cm]\Large\begin{center}{\bf THE SPINLESS
RELATIVISTIC HULTH\'EN PROBLEM}\\[1cm]\large{\bf Wolfgang
LUCHA\footnote[1]{\normalsize\ \emph{E-mail address\/}:
wolfgang.lucha@oeaw.ac.at}}\\[.3cm]\normalsize Institute for High
Energy Physics,\\Austrian Academy of Sciences,\\Nikolsdorfergasse
18, A-1050 Vienna, Austria\\[1cm]\large{\bf Franz
F.~SCH\"OBERL\footnote[2]{\normalsize\ \emph{E-mail address\/}:
franz.schoeberl@univie.ac.at}}\\[.3cm]\normalsize Faculty of
Physics, University of Vienna,\\Boltzmanngasse 5, A-1090 Vienna,
Austria\\[2cm]{\normalsize\bf Abstract}\end{center}\normalsize

\noindent The spinless Salpeter equation can be regarded as the
eigenvalue equation of a Hamiltonian that involves the
relativistic kinetic energy and therefore is, in general, a
nonlocal operator. Accordingly, it is hard to find solutions of
this bound-state equation by exclusively analytic means.
Nevertheless, a lot of tools enables us to constrain the resulting
bound-state spectra rigorously. We illustrate some of these
techniques for the example of the Hulth\'en potential.\vspace{3ex}



\renewcommand{\thefootnote}{\arabic{footnote}}\end{titlepage}

\section{Incentive: Semirelativistic Bound-State Equations}The
spinless Salpeter equation --- encountered in the course of the
nonrelativistic reduction of the Bethe--Salpeter formalism
\cite{BSEa,BSEb,BSEc} for the relativistic description of bound
states within the domain of quantum field theory --- is the
eigenvalue equation of a Hamiltonian $H$~which involves, apart
from some interaction potential $V(\bm{x}),$ the relativistically
correct expression for the kinetic energy $T(\bm{p})$ and, for two
bound-state constituents of equal masses $m,$ reads\footnote{We
present the following discussion in terms of natural units
convenient for particle physics:~$\hbar=c=1.$}\begin{equation}
H\equiv T(\bm{p})+V(\bm{x})\ ,\qquad T(\bm{p})\equiv
2\,\sqrt{\bm{p}^2+m^2}\ ;\label{Eq:H}\end{equation}as such, it
provides a relativistic generalization of the nonrelativistic
Schr\"odinger equation.

We study the spectral features of the Hamiltonian operator
(\ref{Eq:H}) for the particular~case of the interaction potential
$V(\bm{x})$ being the short-range Hulth\'en potential $V_{\rm
H}(r),$ a spherically symmetric central potential
$V(\bm{x})=V(r),$ depending only on the radial coordinate
$r\equiv|\bm{x}|$ and characterized by just two (positive)
parameters, its coupling strength $h$ and its range~$b$:
\begin{equation}V_{\rm H}(r)=-\frac{h}{\exp(b\,r)-1}\ ,\qquad b>0\
,\qquad h\ge0\ .\label{Eq:VH}\end{equation}This potential exhibits
a Coulomb-like singularity at $r=0,$ as is clear from inspection
of its behaviour for small $r,$ and enjoys frequent application in
several different realms~of~physics.

In our analysis, we would like to demonstrate how to take a
rigorous look at the~discrete spectrum of the Hamiltonian $H$ with
Hulth\'en potential (\ref{Eq:VH}) \emph{without\/} having to rely
either on assumptions allowing us to derive \emph{approximate\/}
analytical solutions or on merely numerical approaches. After
discussing, in Sec.~\ref{Sec:EB}, various techniques for deducing
bounds on operator eigenvalues and, in Sec.~\ref{Sec:HP},
Hulth\'en-specific issues, we impose, in Sec.~\ref{Sec:App2}, all
these tools~on~$H.$

\subsection{Nonrelativistic Reduction to Schr\"odinger Hulth\'en
Problem}\label{Sec:HNR}For later use, we recall already here
well-known results on the nonrelativistic (NR)~Hulth\'en problem,
posed by the Schr\"odinger Hamiltonian $H_{\rm NR}$ found as NR
limit of the operator~(\ref{Eq:H}):
\begin{equation}H_{\rm NR}\equiv2\,m+\frac{\bm{p}^2}{m}+V_{\rm
H}(r)\qquad(m_1=m_2=m)\ .\label{Eq:HNR}\end{equation}The generic
one- or two-particle Schr\"odinger Hamiltonian operator with
Hulth\'en potential$$H'\equiv\frac{\bm{p}^2}{2\,\mu}+V_{\rm H}(r)\
,\qquad\mu>0\ ,$$where here and only here $\mu$ indicates either
the mass $m$ of the single bound particle,~$\mu=m,$ or the reduced
mass of the bound two-particle system, $\mu\equiv
m_1\,m_2/(m_1+m_2),$ respectively, possesses a highly welcome
property. For any bound states with orbital angular momentum
quantum number $\ell=0$ (``s waves''), the eigenvalues of $H'$ may
be given in analytic~form~\cite{Flugge}:\begin{equation}
E'_n=-\frac{\left(2\,\mu\,h-n^2\,b^2\right)^2}{8\,\mu\,n^2\,b^2}\
,\qquad n=1,2,3,\dots\ .\label{Eq:F}\end{equation}The parameters
$b$ and $h$ determining the Hulth\'en potential's shape and the
radial quantum number $n$ are subject to a constraint that limits
the number of possible $\ell=0$ bound states:$$n^2\,b^2\le2\,\mu
\,h\qquad\Longleftrightarrow\qquad n\le\frac{\sqrt{2\,\mu\,h}}{b}\
.$$Clearly, the eigenvalues (\ref{Eq:F}) form the $\ell=0$ binding
energies of the bound-state problem~(\ref{Eq:HNR}).

\section{Bounds on Spinless-Salpeter Energy Eigenvalues}
\label{Sec:EB}Before addressing Hulth\'en peculiarities, we recall
standard means of localizing eigenvalues.

\subsection{Upper Energy Bounds}

\subsubsection{Nonrelativistic Kinematics: Schr\"odinger Upper
Bounds on Eigenvalues} \label{Sec:SUB}Due to the concavity of the
square-root operator of the relativistic free energy as a function
of $\bm{p}^2,$ the nonrelativistic limit of that operator forms
the tangent at their point of tangency $\bm{p}^2=0.$ The
implications for the associated Hamiltonians and their eigenvalues
are evident:$$H\equiv T(\bm{p})+V(\bm{x})\le H_{\rm
NR}\equiv2\,m+\frac{\bm{p}^2}{m}+V(\bm{x})\qquad\Longrightarrow\qquad\
E_k\le E_{k,{\rm NR}}\quad\forall\ k=0,1,2,\dots\ .$$

\subsubsection{Relativistic Kinematics: Variational Upper Bounds on
Eigenvalues}\label{Sec:RUB}For arbitrary self-adjoint
Hilbert-space operators $H$ bounded from below, with eigenvalues
$E_k,$ $k=0,1,2,\dots,$ ordered by $E_0\le E_1\le E_2\le\cdots,$
the Rayleigh--Ritz variational method offers a proven instrument
to localize the eigenvalues $E_k$: the $d$ likewise ordered
eigenvalues $\widehat E_k,$ $k=0,1,\dots,d-1,$ of this operator
$H$ restricted to some $d$-dimensional trial subspace of the
domain of $H$ are upper bounds to the lowest-lying $d$ eigenvalues
of $H$ below the onset of its essential spectrum, that is to say,
$E_k\le\widehat E_k$ for all $k=0,1,\dots,d-1.$ It is
straightforward to improve the accuracy
\cite{Lucha:Q&Aa,Lucha:Q&Ab} of these upper bounds by enlarging
the chosen trial subspace. If the basis of this trial subspace is
given analytically in both configuration and momentum space,
finding expectation values of the Hamiltonian (\ref{Eq:H}) may be
considerably facilitated by calculating expectation values of
$T(\bm{p})$ in momentum space and expectation values of
$V(\bm{x})$ in configuration space; we enforce this feature by an
appropriate choice of our basis~vectors. In configuration space,
our orthonormal basis functions $\phi_{k,\ell m}(\bm{x})$ are
defined in terms of the generalized-Laguerre orthogonal
polynomials \cite{AS,Bateman}, $L_k^{(\gamma)}(x)$, for the
parameter $\gamma=2\,\ell+2\,\beta$:\begin{align}&\phi_{k,\ell
m}(\bm{x})=\sqrt{\frac{(2\,\mu)^{2\ell+2\beta+1}\,k!}
{\Gamma(2\,\ell+2\,\beta+k+1)}}\,|\bm{x}|^{\ell+\beta-1}
\exp(-\mu\,|\bm{x}|)\,L_k^{(2\ell+2\beta)}(2\,\mu\,|\bm{x}|)\,
{\cal Y}_{\ell m}(\Omega_{\bm{x}})\ ,\label{Eq:xLag}\\
&L_k^{(\gamma)}(x)\equiv\sum_{t=0}^k\binom{k+\gamma}{k-t}
\frac{(-x)^t}{t!}\ ,\qquad k=0,1,2,\dots\ ,\qquad\mu\in(0,\infty)\
,\qquad\beta\in\!\left(-\frac{1}{2},\infty\right).\nonumber\end{align}
By Fourier transformation, our orthonormal basis functions in
momentum space, $\widetilde\phi_{k,\ell m}(\bm{p}),$ involve the
hypergeometric function $F(u,v;w;z),$ given in terms of the gamma
function \cite{AS}:\begin{align}\widetilde\phi_{k,\ell
m}(\bm{p})&=\sqrt{\frac{(2\,\mu)^{2\ell+2\beta+1}\,k!}
{\Gamma(2\,\ell+2\,\beta+k+1)}}\,\frac{(-{\rm
i})^\ell\,|\bm{p}|^\ell}{2^{\ell+1/2}\,
\Gamma\!\left(\ell+\frac{3}{2}\right)}\nonumber\\[1ex]&\times
\sum_{t=0}^k\,\frac{(-1)^t}{t!}\binom{k+2\,\ell+2\,\beta}{k-t}
\frac{\Gamma(2\,\ell+\beta+t+2)\,(2\,\mu)^t}
{(\bm{p}^2+\mu^2)^{(2\ell+\beta+t+2)/2}}\nonumber\\[1ex]&\times
F\!\left(\frac{2\,\ell+\beta+t+2}{2},-\frac{\beta+t}{2};
\ell+\frac{3}{2};\frac{\bm{p}^2}{\bm{p}^2+\mu^2}\right){\cal
Y}_{\ell m}(\Omega_{\bm{p}})\ ,\label{Eq:pLag}\\&F(u,v;w;z)\equiv
\frac{\Gamma(w)}{\Gamma(u)\,\Gamma(v)}\,\sum_{n=0}^\infty\,
\frac{\Gamma(u+n)\,\Gamma(v+n)}{\Gamma(w+n)}\,\frac{z^n}{n!}\
.\nonumber\end{align}

\subsection{Lower Energy Bounds}\label{Eq:LEB}Lower limits to the
spinless relativistic Hulth\'en problem result from the Coulomb
potential
\begin{equation}V_{\rm C}(r)=-\frac{\kappa}{r}\label{Eq:CP}\
.\end{equation}In the limit $b\downarrow0,$ the Hulth\'en
potential approaches from above the Coulomb-like potential
$$V(r)=-\frac{h}{b\,r}\ .$$Accordingly, the Coulomb potential
(\ref{Eq:CP}) constitutes a lower bound to the Hulth\'en potential
(\ref{Eq:VH}) for sufficiently large Coulomb couplings, more
precisely, for any coupling $\kappa$ that satisfies
$$\kappa\ge\frac{h}{b}\ .$$Precisely the same conclusion follows,
from the series expansion of the exponential $\exp(b\,r)$ in the
denominator of $V_{\rm H}(r),$ Eq.~(\ref{Eq:VH}): $\exp(b\,r)\ge
1+b\,r$. We thus get the operator inequality$$V_{\rm C}(r)\equiv
-\frac{\kappa}{r}\le-\frac{h}{b\,r}\le-\frac{h}{\exp(b\,r)-1}
\equiv V_{\rm H}(r)\qquad\mbox{for}\qquad\frac{h}{b}\le\kappa\ .$$
For the semirelativistic Coulomb bound states, in turn, there
exist well-known lower limits:\begin{itemize}\item In a thorough
mathematical analysis \cite{Herbst} of the spinless relativistic
Coulomb problem, Herbst proved\footnote{We refrain from
explicating here in detail the domains on which the encountered
operators are defined.} that the Hamiltonian (\ref{Eq:H}) with the
Coulomb potential $V_{\rm C}$ is essentially self-adjoint for all
$\kappa\le1,$ that its Friedrichs extension exists up to its
critical coupling$$\kappa_{\rm c}=\frac{4}{\pi}=1.273239\dots\ ,$$
and that, for all $\kappa<\kappa_{\rm c},$ the spectrum
$\sigma(H)$ of the operator $H$ is bounded
from~below:$$\sigma(H)\ge2\,m\,
\sqrt{1-\left(\frac{\kappa}{\kappa_{\rm c}}\right)^{\!2}}=
2\,m\,\sqrt{1-\left(\frac{\pi\,\kappa}{4}\right)^{\!2}}\ .$$\item
Martin and Roy \cite{Martin89} sharpened this lower energy bound
for coupling constants $\kappa\le1$:$$\sigma(H)\ge2\,m\,
\sqrt{\frac{1+\sqrt{1-\kappa^2}}{2}}\ .$$\end{itemize}

\section{Existence and Number of Hulth\'en Bound States}
\label{Sec:HP}\subsection{Semirelativistic vs.\ Nonrelativistic
Number of Bound States}Already in Sec.~\ref{Sec:SUB}, we pointed
out a trivial fact \cite{Lucha96:RCP}: since the nonrelativistic
free~energy,$$T_{\rm NR}(\bm{p})\equiv2\,m+\frac{\bm{p}^2}{m}\
,$$obviously constitutes an upper bound to the corresponding
relativistic kinetic energy $T(\bm{p}),$ a fixed spinless-Salpeter
energy eigenvalue is never larger than its Schr\"odinger
counterpart:$$T(\bm{p})\le T_{\rm
NR}(\bm{p})\qquad\Longrightarrow\qquad H\le H_{\rm
NR}\qquad\Longrightarrow\qquad E_k\le E_{k,{\rm NR}}\ ,\qquad
k=0,1,2,\dots\ .$$Hence, we are led to conclude that the total
number of bound states of the spinless~Salpeter equation, $N,$
will not be less than the number of Schr\"odinger bound states,
$N_{\rm NR}$: $N\ge N_{\rm NR}.$

\subsection{Maximum Number of Nonrelativistic Hulth\'en Bound
States}\label{Sec:BB}The nonrelativistic Hulth\'en problem as
posed by the Schr\"odinger operator (\ref{Eq:HNR}) admits --- in
contrast to the nonrelativistic Coulomb problem --- merely a
finite number of bound states. Bargmann \cite{Bargmann} proved a
simple upper bound to the \emph{total\/} number of NR
bound~states,~$N_{\rm NR}$: $$N_{\rm NR}\lneqq\frac{I\,(I+1)}{2}\
,\qquad I\equiv m\int\limits_0^\infty{\rm d}r\,r\,|V_{\rm
H}(r)|=\frac{\pi^2\,m\,h}{6\,b^2}\ .$$

\subsection{Critical Parameters of the Semirelativistic Hulth\'en
Problem}For a semirelativistic Hamiltonian (\ref{Eq:H}) with
Hulth\'en potential (\ref{Eq:VH}), boundedness from below of this
operator requires that the ratio of coupling strength $h$ over
range parameter $b$ of this potential must not be larger than a
certain critical value of this quotient $h/b.$ That is to say, for
a given value of $b,$ the coupling $h$ must not be larger than its
critical value, whereas, for a given value of $h,$ the range $b$
has to be greater than its critical value. This fact may be easily
demonstrated by application of the Rayleigh--Ritz
variational~technique~(briefly recalled in Sec.~\ref{Sec:RUB}) to
this semirelativistic Hulth\'en problem. To follow as far as
possible~the~analytic path, we try the simplest of the set of
Laguerre basis states in Eqs.~(\ref{Eq:xLag}) or (\ref{Eq:pLag}),
defined by the choices $k=\ell=m=0$ for its quantum numbers and
$\beta=1$ for the variational~parameter~$\beta$:
$$\phi_{0,00}(\bm{x})=\sqrt{\frac{\mu^3}{\pi}}\exp(-\mu\,|\bm{x}|)\
,\qquad\widetilde\phi_{0,00}(\bm{p})=\frac{\sqrt{8\,\mu^5}}{\pi}\,
\frac{1}{(\bm{p}^2+\mu^2)^2}\ .$$For each value of the variational
parameter $\mu,$ \emph{i.e.}, for \emph{all\/} $0<\mu<\infty,$ the
expectation value $\langle H\rangle$ of our Hamiltonian
$H=T(\bm{p})+V_{\rm H}(r)$ with respect to this trial state
provides an upper bound to the ground-state energy. The
expectation value $\langle T(\bm{p})\rangle$ of the kinetic~energy
reads\begin{align*}\langle
T(\bm{p})\rangle&=\frac{4}{3\,\pi\,(m^2-\mu^2)^{5/2}}\\
&\times\left[\mu\,\sqrt{m^2-\mu^2}\,(3\,m^4-4\,m^2\,\mu^2+4\,\mu^4)
+3\,m^4\,(m^2-2\,\mu^2)\,\ArcSec{\frac{m}{\mu}}\right];\end{align*}
the expectation value $\langle V_{\rm H}(r)\rangle$ of the
Hulth\'en potential makes use of a polygamma function,$$\langle
V_{\rm H}(r)\rangle=\frac{4\,h\,\mu^3}{b^3}\,
\psi^{(2)}\!\left(1+\frac{2\,\mu}{b}\right),$$defined \cite{AS},
at some order $n,$ as $(n+1)$-th derivative of the logarithm of
the gamma~function,$$\psi^{(n)}(z)\equiv\frac{{\rm d}^{n+1}}{{\rm
d}z^{n+1}}\ln\Gamma(z)\ ,\qquad\Re z>0\ .$$We are interested in
the limit $\mu\to\infty.$ Expanding $\langle H\rangle=\langle
T(\bm{p}) \rangle+\langle V_{\rm H}(r)\rangle$ for
large~$\mu$~yields$$\langle
H\rangle=\left(\frac{16}{3\,\pi}-\frac{h}{b}\right)\mu+\frac{h}{2}
+\frac{1}{\mu}\left(\frac{16\,m^2}{3\,\pi}-\frac{h\,b}{8}\right)
+O\!\left(\frac{\log\mu}{\mu^3}\right).$$For negative coefficients
of $\mu,$ this expectation value decreases, for rising $\mu,$
without~bound:$$\langle H\rangle\xrightarrow[\mu\to\infty]{}
-\infty\qquad\mbox{for}\qquad\frac{h}{b}>\frac{16}{3\,\pi}\ .$$
Thus, boundedness from below of the operator $T(\bm{p})+V_{\rm
H}(r)$ requires the ratio $h/b$ to~satisfy
\begin{equation}\frac{h}{b}\le\frac{16}{3\,\pi}=1.69765\dots\
.\label{Eq:BB}\end{equation}

\section{Applications}\label{Sec:App2}After all the preparatory
considerations in Secs.~\ref{Sec:EB} and \ref{Sec:HP}, it is now
rather straightforward to apply the insights gained thereby to the
semirelativistic Hulth\'en Hamiltonian under study. For ease of
comparison, we would like to do this exercise, of course, for a
choice of numerical values of the bound-state components' mass $m$
and the Hulth\'en potential parameters $b$ and $h$ that has been
also adopted in, at least, one previous investigation of the
present problem. We are aware of merely two publications
\cite{IS_IJMPE08,Zarrinkamar} discussing spinless Salpeter
equations with either the original \cite{Zarrinkamar} form
(\ref{Eq:VH}) of Hulth\'en's potential or a properly generalized
\cite{IS_IJMPE08}~variant thereof. Both of these works rely on
various simplifying modifications in order to arrive at a
Schr\"odinger-like implicit eigenvalue equation that is assumed to
represent some reasonable approximation to the spinless Salpeter
equation but allows for obtaining analytic solutions.
Unfortunately, only Ref.~\cite{Zarrinkamar} illustrates its
resulting expressions by explicit examples; from Table 1 therein,
for use as parameter values in what follows, we read off, in
arbitrary units,\footnote{Repeated inspection of the definition of
the Hulth\'en potential provided by Eq.~(6) of
Ref.~\cite{Zarrinkamar} prompts us to take the strange minus sign
in the caption of Table 1 of Ref.~\cite{Zarrinkamar} not too
literally; moreover, we~do~not wonder about the meaning of the
parameter $h=1,$ mentioned in the caption of this table but
nowhere~else.}\begin{equation}m=1\ ,\qquad b=0.15\ ,\qquad h=0.11\
.\label{Eq:Z+P}\end{equation}The reliability of these approximate
solutions may be immediately checked by our findings:
\begin{itemize} \item First of all, the parameter values
(\ref{Eq:Z+P}) satisfy the inequality (\ref{Eq:BB}) imposed by
demanding the spinless-Salpeter Hamiltonian with Hulth\'en
potential to be bounded from below:
$$\frac{h}{b}=0.7\dot3<\frac{16}{3\,\pi}=1.69765\dots\ .$$Thus,
for the setting (\ref{Eq:Z+P}) one may expect, on good grounds, to
find bound states~at~all.\item According to the inequality
limiting the quantum number $n$ of s-wave bound states in
Sec.~\ref{Sec:HNR}, the choice (\ref{Eq:Z+P}) allows for just two
nonrelativistic $\ell=0$ Hulth\'en bound~states:
$$n\le2<\frac{\sqrt{m\,h}}{b}=2.211\dots\ .$$\item The Bargmann
bound of Sec.~\ref{Sec:BB} shows that the nonrelativistic
Hulth\'en problem can accommodate at most 36 bound states since,
upon use of the values (\ref{Eq:Z+P}), it returns, for the total
number of bound states, $N_{\rm NR}<36.357\dots,$ which is,
potentially, still rather far from optimum. Improvements of the
Bargmann bound exist copiously but usually lead to expressions
that are much harder to deal with than Bargmann's~simple result.
\item In order to maximize our lower bound to the spectrum of the
semirelativistic Hulth\'en problem resulting from the observation
made in Sec.~\ref{Eq:LEB} that for $\kappa\ge h/b$ the Hulth\'en
potential is bounded from below by the Coulomb potential, we
present this bound for the minimum possible value of the Coulomb
coupling that still guarantees the desired operator inequality,
viz., $\kappa=h/b=0.7\dot3,$ for which both Coulomb lower limits
apply. This yields, for the ground-state energy eigenvalue $E_0$
and the corresponding~binding energy $B_0\equiv E_0-2\,m,$ from
the Herbst lower bound \cite{Herbst} $E_0\ge1.635$ and
$B_0\ge-0.365,$ respectively, and from its Martin--Roy counterpart
\cite{Martin89} $E_0\ge1.833$ and $B_0\ge-0.167,$ respectively.
Surprisingly, one entry in Table 1 of Ref.~\cite{Zarrinkamar}
slightly violates this result.\item Table \ref{Tab:RHP} summarizes
upper limits on the lowest-lying bound-state levels of the
spinless relativistic Hulth\'en problem found along the lines
sketched in Secs.~\ref{Sec:SUB} and \ref{Sec:RUB} by variational
approach or standard numerical solution of the Schr\"odinger
equation \cite{Lucha98}.

\begin{table}[ht]\caption{Upper limits to the binding energy for
the lowest-lying bound states of the spinless Salpeter equation
with Hulth\'en's potential, for the parameter values of
Ref.~\cite{Zarrinkamar}: the trivial Schr\"odinger bounds
$\overline{E}_{\rm NR}$ of Subsec.~\ref{Sec:SUB} and the Laguerre
bounds $\overline{E}$ of Subsec.~\ref{Sec:RUB}. Any bound state is
identified by its radial quantum number, $n_r,$ and orbital
angular~momentum quantum number, $\ell$. Merely for illustration,
we keep the dimension $d$ of the variational trial space $D_d$ and
both variational parameters $\mu$ and $\beta$ fixed to the values
$d=25,$~$\mu=1,$~$\beta=1.$}\label{Tab:RHP}
\begin{center}\begin{tabular}{ccll}\hline\hline\\[-1.5ex]
\multicolumn{2}{c}{Bound state}&\multicolumn{1}{c}{Spinless
Salpeter equation}&\multicolumn{1}{c}{Schr\"odinger equation}
\\[1ex]$n_r$&$\ell$&\multicolumn{1}{c}{$\overline{E}(n_r,\ell)$}&
\multicolumn{1}{c}{$\overline{E}_{\rm NR}(n_r,\ell)$}\\[1.5ex]
\hline\\[-1.5ex]0&0&$\qquad\quad-0.10577$&$\qquad-0.085069\dot4$
\\[1ex]1&0&$\qquad\quad-0.0022398$&$\qquad-0.00\dot1$\\[1.5ex]
\hline\hline\end{tabular}\end{center}\end{table}

\item In order to approximate the spinless Salpeter equation by an
equation easier to treat, assuming the bound-state constituents to
be sufficiently heavy both of the two earlier investigations
mentioned above \cite{IS_IJMPE08,Zarrinkamar} prefer to expand
that cumbersome relativistic kinetic energy $T(\bm{p})$ in our
semirelativistic Hamiltonian $H$ nonrelativistically and, by
retaining terms up to order $\bm{p}^4/m^4,$ to get some pseudo
spinless-Salpeter~Hamiltonian$$H_{\rm p}\equiv2\,m
+\frac{\bm{p}^2}{m}-\frac{\bm{p}^4}{4\,m^3}+V(\bm{x})$$that is
obviously unbounded from below \cite{Lucha14:SRWSP}, so that the
term $\bm{p}^4/(4\,m^3)$ can be taken into account only
perturbatively. Anyway, we may check whether such nonrelativistic
expansion is justifiable at all, by assuming that the lowest bound
state emerging~from our variational procedure provides a
satisfactory description of the ground state, and by inspecting,
for this state, the expectation value of the next-to-lowest
term~in~$T(\bm{p})$:$$\left\langle\frac{\bm{p}^2}{m^2}\right\rangle
\approx0.26\ ;$$thus, the system governed by the parameters
(\ref{Eq:Z+P}) can be viewed as not too relativistic.\end{itemize}

\section{Summary and Concluding Remarks}The spinless Salpeter
equation forms the penultimate stage in the nonrelativistic
reduction of the homogeneous Bethe--Salpeter equation (more
details than those presented above~can be found in, \emph{e.g.},
Refs.~\cite{Lucha91, Lucha94:Como,Lucha99:talks,Lucha04:TWR,
Lucha05:IBSEWEP}). Perhaps because of the paramount importance of
its~origin or because of the challenge represented by the nonlocal
nature of the Hamiltonian operator controlling the bound states
under study, we can witness, from time to time, a considerable
increase in interest in this equation of motion, which, in turn,
motivated the above analysis aiming at the discussion of a couple
of rigorous constraints on the spectrum of bound states to be
expected if in our spinless Salpeter equation all interactions
between the bound-state constituents are subsumed by a Hulth\'en
potential. Needless to say, singular potentials such as that
introduced by Hulth\'en pose obstacles which differ from those to
be faced if studying non-singular interactions such as the
Woods--Saxon potential \cite{Lucha14:SRWSP}. Anyway, a solid
starting point for studying semirelativistic systems is the
corresponding~nonrelativistic case \cite{Varshni,Jia_IJMPA09}.


\small\begin{thebibliography}{30}
\bibitem{BSEa}H.~A.~Bethe and E.~E.~Salpeter, Phys.~Rev.~{\bf
82} (1951) 309.
\bibitem{BSEb}M.~Gell-Mann and F.~Low,~Phys.~Rev.~{\bf 84} (1951)
350.
\bibitem{BSEc}E.~E.~Salpeter and H.~A.~Bethe, Phys.~Rev.~{\bf 84}
(1951) 1232.
\bibitem{Flugge}S.~Fl\"ugge, \emph{Practical Quantum Mechanics\/}
(Springer, Berlin, 1994), Vol.~I, pp.~175--178.
\bibitem{Lucha:Q&Aa}W.~Lucha and F.~F.~Sch\"oberl, Phys.~Rev.~A {\bf
60} (1999) 5091, arXiv:hep-ph/9904391.
\bibitem{Lucha:Q&Ab}W.~Lucha and F.~F.~Sch\"oberl,
Int.~J.~Mod.~Phys.~A {\bf 15} (2000) 3221, arXiv:hep-ph/9909451.
\bibitem{AS}M.~Abramowitz and I.~A.~Stegun (eds.), \emph{Handbook
of Mathematical Functions\/} (Dover, New York, 1964).
\bibitem{Bateman}Bateman Manuscript Project, A.~Erd\'elyi \emph{et
al.}, \emph{Higher Transcendental Functions} (McGraw--Hill, New
York, 1953), Vol.~II.
\bibitem{Herbst}I.~W.~Herbst, Commun.~Math.~Phys.~{\bf 53} (1977)
285; {\bf 55} (1977) 316 (addendum).
\bibitem{Martin89}A.~Martin and S.~M.~Roy, Phys.~Lett.~B {\bf 233}
(1989) 407.
\bibitem{Lucha96:RCP}W.~Lucha and F.~F.~Sch\"oberl,
Phys.~Rev.~A {\bf 54} (1996) 3790, arXiv:hep-ph/9603429.
\bibitem{Bargmann}V.~Bargmann, Proc.~Natl.~Acad.~Sci.~USA {\bf
38} (1952) 961.
\bibitem{IS_IJMPE08}S.~M.~Ikhdair and R.~Sever,
Int.~J.~Mod.~Phys.~E {\bf 17} (2008) 1107.
\bibitem{Zarrinkamar}S.~Zarrinkamar, A.~A.~Rajabi, H.~Hassanabadi,
and H.~Rahimov, Phys.~Scr.~{\bf 84} (2011) 065008.
\bibitem{Lucha98}W.~Lucha and F.~F.~Sch\"oberl,
Int.~J.~Mod.~Phys.~C {\bf 10} (1999) 607, arXiv:hep-ph/9811453.
\bibitem{Lucha14:SRWSP}W.~Lucha and F.~F.~Sch\"oberl,
Int.~J.~Mod.~Phys.~A {\bf 29} (2014) 1450057,
arXiv:1401.5970~[hep-ph].
\bibitem{Lucha91}W.~Lucha, F.~F.~Sch\"oberl, and D.~Gromes,
Phys.~Rep.~{\bf 200} (1991) 127.
\bibitem{Lucha94:Como}W.~Lucha and F.~F.~Sch\"oberl, in
\emph{Proceedings of the International Conference on Quark
Confinement and the Hadron Spectrum\/}, edited by N.~Brambilla and
G.~M.~Prosperi (World Scientific, River Edge, NJ, 1995) p.~100,
arXiv:hep-ph/9410221.
\bibitem{Lucha99:talks}W.~Lucha and F.~F.~Sch\"oberl,
Int.~J.~Mod.~Phys.~A {\bf 14} (1999) 2309, arXiv:hep-ph/9812368.
\bibitem{Lucha04:TWR}W.~Lucha and F.~F.~Sch\"oberl, Recent
Res.~Dev.~Phys.~{\bf 5} (2004) 1423, arXiv:hep-ph/0408184.
\bibitem{Lucha05:IBSEWEP}W.~Lucha and F.~F.~Sch\"oberl, J.~Phys.~G
{\bf 31} (2005) 1133, arXiv:hep-th/0507281.
\bibitem{Varshni}Y.~P.~Varshni, Phys.~Rev.~A {\bf 41} (1990) 4682.
\bibitem{Jia_IJMPA09}C.-S.~Jia, Y.-F.~Diao, L.-Z.~Yi, and T.~Chen,
Int.~J.~Mod.~Phys.~A {\bf 24} (2009) 4519.
\end{thebibliography}
\end{document}